\documentclass[pra,twocolumn,showpacs,superscriptaddress,10pt]{revtex4-1} %
\usepackage{amsmath,amssymb,graphicx}
\usepackage{subfigure}
\usepackage{color}
\usepackage{placeins}

\usepackage{pgfplots}
\usepackage{tikz}
\usetikzlibrary{arrows,decorations.markings}

\begin{document}

\title{Space-time duality and quantum temporal imaging}
\author{Giuseppe Patera}
\affiliation{Univ.~Lille, CNRS, UMR 8523 - PhLAM - Physique des Lasers Atomes et Mol\'{e}cules, F-59000 Lille, France}
\author{Dmitri B.~Horoshko}
\affiliation{Univ.~Lille, CNRS, UMR 8523 - PhLAM - Physique des Lasers Atomes et Mol\'{e}cules, F-59000 Lille, France}
\affiliation{B. I. Stepanov Institute of Physics, NASB, Nezavisimosti Avenue 68, Minsk 220072, Belarus}
\author{Mikhail I.~Kolobov}
\affiliation{Univ.~Lille, CNRS, UMR 8523 - PhLAM - Physique des Lasers Atomes et Mol\'{e}cules, F-59000 Lille, France}

\begin{abstract}
Using the space-time analogy, we compare the performance of quantum temporal imaging with its classical counterpart.
We consider a temporal imaging scheme, based on the sum-frequency generation (SFG) time lens, but our results can be applied
to other temporal imaging schemes such as, for instance, four-wave mixing. Extending the theory presented in our
previous publications, in this paper we take into account the finite time aperture of the imaging system, characterized
by its pupil function. Using the quantum theory, we obtain a unitary transformation of the quantum field from the input
to the output of the imaging scheme and identify the contribution of the vacuum fluctuations missing in the classical
theory. This contribution plays a key role in the quantum temporal imaging of nonclassical temporal waveforms,
characterized by nonclassical fluctuations of the electromagnetic field. As an example, we consider quantum temporal
imaging of broadband squeezed light and formulate the criteria for conservation of its squeezing properties at the
output of the system.
\end{abstract}
\date{\today}
\maketitle

\section{Introduction}

Optical temporal imaging is a technique for manipulating temporal waveforms similar to manipulating
spatial transverse wave-surfaces using a space-time analogy~\cite{Company2011,Salem2013}. Temporal
imaging was first discovered in the mid-1960s with purely electrical systems, then extended to optics \cite{Akhmanov1969}, and later converted into
all-optical technology using the development of nonlinear optics and ultra-short-pulse lasers~\cite{Kolner1994}.
Nowadays, temporal imaging has become a mature area of modern optics with various applications. One of the
typical applications of temporal imaging is stretching of ultra-fast temporal waveforms with bandwidths of
tens and hundreds of THz by several orders of magnitude in order to make them detectable by ordinary
photodetectors with bandwidths of the order of tens of GHz. Another example is compression of waveforms
created by electro-optical devices to picosecond and sub-picosecond timescales for increasing the rate of
their transmission.

The key element of a temporal imaging system is a time lens, which introduces a quadratic temporal phase
modulation into an input waveform, similarly to quadratic phase modulation in the transverse dimension, introduced
by its spatial counterpart into an input wave-surface. Optical time lenses presently are based on electro-optical phase modulation \cite{Giordmaine1968,Grischowsky1974,Kolner1988,Kolner1989,Karpinski2016}, sum-frequency generation (SFG) \cite{Agrawal1989,Bennett1994,Bennett1999,Bennett2000a,Bennett2000b,Bennett2001,Hernandez2013}, or four-wave mixing \cite{Foster2008,Foster2009,Okawachi2009,Kuzucu2009} and provide a temporal magnification up to 100 times.

The classical theory of temporal imaging considers the electromagnetic field in a framework of classical electrodynamics.
At the same time, temporal imaging has many potential applications in quantum optics and quantum information, and could
lead to establishment of the whole new branch, {\it quantum temporal imaging}, similar to its spatial analog - quantum imaging  \cite{Lugiato2002,Shih2007,Kolobov2007}.
Quantum temporal imaging should allow for manipulation of nonclassical temporal waveforms in a noiseless fashion, i.~e.~without
destruction of their nonclassical properties such as squeezing, entanglement, or nonclassical photon statistics. Such a technique may find numerous applications in optical implementations of quantum information protocols with discrete and continuous variables. In particular, quantum temporal stretching provides a method of decoding quantum information, conveyed by ultrabroadband squeezed light, generated in chirped quasi-phase-matched crystals \cite{Horoshko2013,Horoshko2017,Chekhova2018}. The primary
goal of the theory of quantum temporal imaging is to establish the physical conditions for such a noiseless performance of temporal imaging
devices. Several papers have addressed the subject of temporal imaging at the single-photon level~\cite{Kielpinski2011,Lavoie2013,Zhu2013,Donohue2015}. In our previous publications~\cite{Patera2015,Patera2017,Shi2017}
we have considered quantum temporal imaging with broadband squeezed light and have formulated the appropriate conditions
for the imaging scheme and the parameters of the light source, which allow for maintaining the squeezing at the output of the scheme.

In the present paper we are extending the quantum theory of temporal imaging, formulated in our previous publications, in order to
take into account the finite time aperture of the imaging system, characterized by its pupil function. In classical temporal imaging
the role of the finite time aperture was investigated in Ref.~\cite{Kolner1994}. Precisely, it was demonstrated that, similarly to
conventional spatial imaging, a finite time aperture determines the resolution of the temporal imaging system. Indeed, the finite
size of the pupil function in a temporal imaging system imposes the upper limit of temporal frequencies that can be transmitted
through the system. The frequencies above this limit are lost and, therefore, are not present in the image at its output. In
quantum theory all losses are accompanied by the fluctuations in order to preserve the unitarity of the field transformation.
Therefore, the primary task of the quantum temporal imaging theory is to find out a unitary transformation of the field from
the input to the output of the imaging system with the finite pupil, and determine the contribution of the quantum fluctuations
missing in the classical theory.
In this paper we present such unitary transformation described by two corresponding impulse response functions. The first one is identical to the impulse response function in the classical temporal imaging, while the second one is introduced in this paper for the first time and is absent in the classical theory. This impulse response function describes quantum temporal imaging of the quantum vacuum fluctuations always present at the input of the imaging scheme.

The paper is organized as follows. In Sect.~II we formulate the quantum theory of temporal imaging with an SFG time lens and a finite
time window. In Sect.~III we make a detailed comparison between the quantum theory and its classical counterpart and explain the
difference between them. In Sect.~IV we apply our quantum theory to quantum temporal imaging with nonclassical light, taking as an
example a broadband squeezed light. In Sect.~V we provide a conclusion and give an outlook for the future.

\section{Quantum temporal imaging with a SFG time lens}

\subsection{Description of the scheme}

We consider a simple temporal imaging system shown in Fig.~1. It consists of the first dispersive medium followed by a time lens and the second dispersive medium. In the following we will refer to the first (second) medium as the \textit{input} (\textit{output}) dispersive medium. The time lens is implemented by a nonlinear process that can be either a sum-frequency generation (SFG) or a four-wave mixing (FWM). In this paper we consider the SFG time lens. The case of the FWM time lens was considered in Refs.~\cite{Shi2017,Shi2018}. In the SFG process a strong pump wave of the frequency $\omega_{p}$ interacts with a signal wave of the frequency $\omega_{s}$ to produce an idler wave of the frequency $\omega_{i}$ such that $\omega_{s}+\omega_{p}=\omega_{i}$.
\begin{figure}[h!]
    \centering
    \includegraphics[width=\columnwidth]{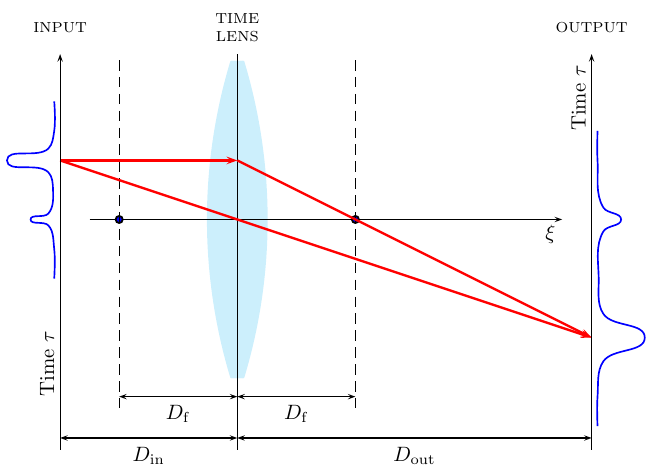}
  \caption{Temporal imaging system with a single time lens. $\xi$ is the propagation distance measured in total group delay dispersion from the object plane, $\tau$ is the local time relative to the group delay. Red arrows are temporal rays \cite{Bennett1999,Bennett2000a} corresponding to different spectral components of the input signal. Image formation is completely analogous to that of a spatial imaging system.}
  \label{fig:TimeLens}
\end{figure}

We use the plane-wave approximation and describe each field by its positive-frequency amplitude ${E}^{(+)}_{\mu}(t,z)$ at the time $t$ and the longitudinal position $z$, where the index $\mu=\{s,i,p\}$ identifies the signal, the idler or the pump waves, respectively. We consider the signal and the idler waves as quantum-mechanical operators and the pump wave as a classical c-function. All waves are assumed to be narrow-band with the carrier frequencies $\omega_{\mu}$. Each wave passing through a medium, experiences dispersion, characterized by the dependence of its wave vector $k_{\mu}(\omega)$ on the frequency $\omega$, which we decompose around the carrier frequency in $\Omega=\omega-\omega_{\mu}$ and limit the Taylor series to the first three terms:
\begin{equation}\label{quadratic}
     k_{\mu}(\omega) \approx k_{\mu}(\omega_{\mu})+\beta_{\mu}^{(1)}\Omega + \beta_{\mu}^{(2)}\Omega^2/2,
\end{equation}
where $\beta_{\mu}^{(1)} = (\mathrm{d} k_{\mu}/\mathrm{d} \Omega)_{\omega_{\mu}}$ is the inverse group velocity, and $\beta_{\mu}^{(2)} = (\mathrm{d}^2 k_{\mu}/\mathrm{d} \Omega^2)_{\omega_{\mu}}$ is the group velocity dispersion of the medium at the carrier frequency $\omega_{\mu}$.

We introduce a frame of reference travelling with the wave at the group velocity, possibly different in each medium. Thus, for each point $z$ we introduce the delayed time $\tau=t-\tau_{\mu}(z)$, where $\tau_{\mu}(z)$ is the total delay for the wave ${E}^{(+)}_{\mu}(t,z)$ from the object plane at $z=z_{\mathrm{in}}$ to the point $z$, between $z_{\mathrm{in}}$ and the image plane $z=z_{\mathrm{out}}$. A delay in a medium of length $L$ with the inverse group velocity $\beta_{\mu}^{(1)}$ is $\beta_{\mu}^{(1)}L$, and the total delay $\tau_{\mu}(z)$ can be found by summing the delays of all media between $z_{\mathrm{in}}$ and $z$.
In this reference frame we can write
\begin{equation}\label{E}
     \hat{E}^{(+)}_{\mu}(t,z) = \mathcal{E}_{\mu} \mathrm{e}^{i\left(k_{\mu} z-\omega_{\mu} t\right)} \hat{A}_{\mu}(\tau,z),
\end{equation}
where $\mathcal{E}_{\mu}$ is the single-photon field amplitude, $k_{\mu}(\omega_{\mu})=k_{\mu}$, and field envelope ${A}_{\mu}(\tau,z)$ is given by
\begin{equation}\label{A}
     \hat{A}_{\mu}(\tau,z) = \frac1{2\pi} \int_{-\infty}^{\infty} \hat{\epsilon}_{\mu}(\Omega,z) \mathrm{e}^{i\beta_{\mu}^{(2)}\Omega^2(z-z_0)/2-i\Omega \tau} \mathrm{d}\Omega,
\end{equation}
with $\hat{\epsilon}_{\mu}(\Omega,z)$ being the slowly-varying quantum amplitude~\cite{Boyd,Kolobov1999} for the given medium with the entrance point at $z_0$.

In a linear dispersive medium the slowly-varying amplitude $\hat{\epsilon}_{\mu}(\Omega,z)$ does not change during the propagation. Hence, the envelope field of $\hat{A}_{\mu}(\tau,z)$ at the end of the input dispersive medium, $z_1$, can be written as,
\begin{equation}\label{disp evo II}
     \hat{A}_{\mu}(\tau,z_1) = \int_{-\infty}^{\infty} G_\mathrm{in}(\tau-\tau')\hat{A}_{\mu}(\tau',z_{\mathrm{in}})\mathrm{d}\tau',
\end{equation}
where
\begin{equation}\label{G}
G_\mathrm{in}(\tau) = \frac{\mathrm{e}^{-i\tau^2/2 D_\mathrm{in}}}{\sqrt{-2\pi i D_\mathrm{in}}}.
\end{equation}
with $D_\mathrm{in}=\beta_{s}^{(2)}(z_1-z_{\mathrm{in}})$ being the group delay dispersion (GDD) of the medium. Equation \eqref{disp evo II} describes a unitary transformation of the field operators, which follows from the identity
\begin{equation}\label{GG*}
 \int_{-\infty}^{\infty} G_\mathrm{in}(\tau-\tau'')G_\mathrm{in}^*(\tau'-\tau'')\mathrm{d}\tau'' = \delta(\tau-\tau'),
\end{equation}
and hence it preserves the commutator for the field operators, as should be for a dispersive propagation without losses.

Assuming the perfect phase matching and the undepleted pump, the SFG time lens can be described by the following unitary transformation from the point $z_1$ (time lens input) to the point $z_2$ (time lens output) \cite{Patera2015},
\begin{eqnarray}\label{as out}
     \hat{A}_{s}(\tau,z_2) &=& c(\tau)\,\hat{A}_{s}(\tau,z_1)-
     s(\tau)\mathrm{e}^{-i\phi(\tau)}\,\hat{A}_{i}(\tau,z_1),
     \\\nonumber
     \hat{A}_{i}(\tau,z_2) &=& s(\tau)\mathrm{e}^{i\phi(\tau)} \hat{A}_{s}(\tau,z_1) + c(\tau)\hat{A}_{i}(\tau,z_1),
\end{eqnarray}
with
\begin{align}\label{u}
     c(\tau)&=\mathrm{cos}(g A_{p}(\tau) L),
     \\\nonumber
     s(\tau)&=\mathrm{sin}(g A_{p}(\tau) L).
\end{align}
Here $L=z_2-z_1$ is the length of the nonlinear medium, while $A_{p}(\tau)$ and $\phi(\tau)$ are the modulus and the phase of the pump pulse. For the implementation of a time lens, a short Gaussian pulse of duration $\tau_{\mathrm{p}}$ is propagated through a dispersive medium of length $L_{p}$ and group velocity dispersion $\beta_{p}^{(2)}$ at the carrier frequency $\omega_{p}$. At the output of the medium the pump pulse is stretched to the duration $T \gg \tau_{p}$ and acquires a phase that is quadratic in time, $\phi(\tau)=\tau^2/2 D_{\mathrm{f}}$, with $D_{\mathrm{f}} = -\beta_{p}^{(2)} L_{\mathrm{p}}$ known as the focal GDD \cite{Bennett2000a,Bennett2000b}. In the present work we consider only the case of negatively chirped pump, $\beta_{p}^{(2)}<0$, for definiteness. As consequence, $D_{\mathrm{f}}>0$. We also assume that the signal and the idler beams pass through the media with positive dispersion, so that both $D_{\mathrm{in}}$ and $D_{\mathrm{out}}$ are positive.

Equations \eqref{as out} describe a unitary transformation of the photon annihilation operators of the signal and the idler waves from the input of the SFG crystal to its output preserving the canonical commutation relations. As follows from their definitions, the coefficients $c(\tau)$ and $s(\tau)$ satisfy the condition
\begin{align}
     |c(\tau)|^2+|s(\tau)|^2=1.
\label{unit}
\end{align}
Therefore, they can be interpreted as the reflection and the transmission coefficients of an equivalent beam splitter. For time lens applications the signal port is injected with an input state while the input idler port is empty. As a consequence, the vacuum fluctuations enter into the process through this port and mix with the input state. Since these vacuum fluctuations are detrimental for the non-classical input states, they need to be avoided. They can be eliminated by setting experimental conditions such that the \textit{conversion efficiency} $|s(\tau)|^2=1$. This condition can be obtained by requiring $g A_{p}(\tau) L=\pi/2$. However, since the pump pulses have a finite duration, the previous conditions cannot be satisfied for all $\tau$. The consequence is that the time lens presents a finite temporal aperture that lets in the vacuum fluctuations.

The transformation of the envelope field $\hat{A}_{i}(\tau,z)$ in the output dispersive medium is given by an equation analogous to that of the input medium,
\begin{equation}\label{disp evo out}
     \hat{A}_{i}(\tau,z_{\mathrm{out}}) = \int_{-\infty}^{\infty} G_\mathrm{out}(\tau-\tau')\hat{A}_{i}(\tau',z_{2})\mathrm{d}\tau',
\end{equation}
where
\begin{equation}\label{G2}
     G_\mathrm{out}(\tau) = \frac{\mathrm{e}^{-i\tau^2/2 D_\mathrm{out}}}{\sqrt{-2\pi i D_\mathrm{out}}}.
\end{equation}
with $D_\mathrm{out}=\beta_{i}^{(2)}(z_{\mathrm{out}}-z_{2})$ being the GDD of the output medium.

\subsection{Impulse response functions for quantum temporal imaging}

Combining Eqs.~(\ref{disp evo II}), (\ref{as out}) and (\ref{disp evo out}) we obtain the transformation of the quantum field operators from the object to the image plane,
\begin{eqnarray}\label{Arect}
    \hat{A}_{\mathrm{out}}(\tau) &=& \frac{i}{\sqrt{|M|}}\exp\left(-\frac{i\tau^2}{2|M|D_\mathrm{f}}\right)\\\nonumber
    &\times& \left\{\int_{-\infty}^{\infty} \tilde{p}\left(\tau,\tau'\right) \hat{A}_\mathrm{in}\left(\frac{\tau'}M\right)d\tau'\right.\\\nonumber
    &+& \left.\int_{-\infty}^{\infty} \tilde{q}\left(\tau,\tau'\right) \hat{B}_\mathrm{in}\left(\frac{\tau'}M\right)d\tau'\right\},
\end{eqnarray}
where we have denoted $\hat{A}_{\mathrm{out}}(\tau)=\hat{A}_i(\tau,z_{\mathrm{out}})$, $\hat{A}_{\mathrm{in}}(\tau)=\hat{A}_s(\tau,z_{\mathrm{in}})$ and $\hat{B}_\mathrm{in}(\tau)=\hat{A}_i(\tau,z_{\mathrm{in}})$. The last operator describes the vacuum field of the idler in the object plane and is absent in the classical temporal imaging theory. The impulse response functions $\tilde{p}(\tau,\tau')$ and $\tilde{q}(\tau,\tau')$ in Eq.~(\ref{Arect}) have the following forms,
\begin{eqnarray}\label{p}
   \tilde{p}(\tau,\tau') &=&  p(\tau-\tau')e^{i\theta(\tau,\tau')},\\\label{q}
   \tilde{q}(\tau,\tau') &=& q(\tau-\tau')e^{i\theta(\tau,\tau')},
\end{eqnarray}
where the function $p(\tau)$ is the point-spread function of the classical imaging transformation~\cite{Kolner1994},
\begin{equation}
   p(\tau)=\frac{1}{2\pi}\int_{-\infty}^{\infty}\mathrm{d}\Omega \mathrm{e}^{\mathrm{i}\tau \Omega} s(D_\mathrm{out}\Omega),
   \label{psf}
\end{equation}
and the function $q(\tau)$ is the second point-spread function necessary for quantum description of our temporal imaging scheme. It describes the temporal imaging of the quantum fluctuations of the field $\hat{B}_\mathrm{in}(\tau)$ and is absent in the classical theory of temporal imaging because such fluctuations do not exist in the classical theory. This point-spread function is given by the following Fourier transform of the coefficient $c(\tau)$ from Eq.~(\ref{u}), properly scaled and phase-adjusted as follows,
\begin{equation}
   q(\tau)=\frac{1}{2\pi}\int_{-\infty}^{\infty}\mathrm{d}\Omega \mathrm{e}^{\mathrm{i}\tau \Omega} c'(D_\mathrm{out}\Omega),
   \label{qpsf}
\end{equation}
with $c'(\tau) = c(\tau)\exp\{-i\tau^2/2D_\mathrm{f}\}$. The phase appearing in Eqs.~\eqref{p} and \eqref{q} is defined as
\begin{eqnarray}\label{theta}
    \theta(\tau,\tau') = \frac{\tau^2-\tau'^2}{2|M|D_\mathrm{out}},
\end{eqnarray}
and in their derivation we have applied the time lens equation \cite{Kolner1994}
\begin{equation}\label{lenseq}
\frac1{D_\mathrm{in}}+\frac1{D_\mathrm{out}} = \frac1{D_\mathrm{f}}
\end{equation}
and the definition of the magnification $M=-D_\mathrm{out}/D_\mathrm{in}$.

The impulse response functions $\tilde{p}(\tau,\tau')$ and $\tilde{q}(\tau,\tau')$ satisfy the relation
\begin{eqnarray}\label{unitarity}
     \int_{-\infty}^{\infty} \tilde{p}(\tau,s)\tilde{p}^*(\tau',s)ds &+& \int_{-\infty}^{\infty} \tilde{q}(\tau,s)\tilde{q}^*(\tau',s)ds \\\nonumber
&=& \delta(\tau-\tau'),
\end{eqnarray}
required by the unitarity of Eq.~(\ref{Arect}).

While the impulse response function $\tilde p(\tau,\tau')$ is well known in the classical temporal imaging theory \cite{Kolner1994,Bennett2000a,Bennett2000b}, the second function $\tilde q(\tau,\tau')$ has not been yet considered to the best of our knowledge. Introduction of this function in the quantum theory together with the corresponding operator $\hat{B}_\mathrm{in}(\tau)$ in Eq.~(\ref{Arect}) is necessary for conservation of the canonical commutation relations of the field operators and the unitarity of the transformation from the input to the output plane.

The impulse response functions $\tilde{p}(\tau,\tau')$ and $\tilde{q}(\tau,\tau')$ in Eqs.~(\ref{p}) and (\ref{q}) are the products of two factors. The first factor is given by the time-invariant point spread functions $p(\tau-\tau')$ and $q(\tau-\tau')$, while the second is a time-variant phase factor. Below we shall formulate the conditions when one can neglect the second factor and use the time invariant approximation for the impulse response functions. Our analysis follows closely the equivalent approximation in the classical spatial imaging~\cite{Goodman2005, Tichenor1972}. We start with the classical field transformation, given by the quantum averages. We show that in the classical temporal imaging the phase $\theta(\tau,\tau')$ can be neglected if the object field is restricted to an interval of duration $T_0$ much shorter than the time lens aperture $T$, $T_0\ll T$.  Indeed, the classical transformation can be obtained by quantum averaging of Eq.~\eqref{Arect} and taking into account that the quantum field $\hat{B}_\mathrm{in}(\tau)$ is in the vacuum state, $\langle \hat{B}_\mathrm{in}\left(\tau\right)\rangle=0$. On the one hand, if we assume that the modulus of the pump pulse is an even function of time (which is a typical experimental situation) of duration $T$, then the point-spread function $p(\tau)$ is a real even function of width approximately $2\pi D_\mathrm{out}/T$. It means that the integrand of the right hand side of Eq.~\eqref{Arect} is substantially non-zero only for $|\tau-\tau'|\le\pi D_\mathrm{out}/T$. On the other hand, if the object is restricted to $|\tau'|\le T_0/2$, then
\begin{equation}
   |\tau+\tau'|\le2|\tau'|+|\tau-\tau'|\le|M|T_0+\pi D_\mathrm{out}/T.
\end{equation}
Combining these two conditions we obtain from Eq.~\eqref{theta} for the area where the integrand is non-zero:
\begin{eqnarray}\label{theta2}
  |\theta(\tau,\tau')| \le \frac{\pi T_0}{2T} + \frac{\pi^2D_\mathrm{out}}{2T^2|M|}.
\end{eqnarray}
The phase $\theta(\tau,\tau')$ can be neglected if in the considered area it is small compared to $\pi/2$, which requires
\begin{eqnarray}\label{theta3}
 \frac{T_0}{T} + \frac{\pi D_\mathrm{out}}{T^2|M|} \ll 1,
\end{eqnarray}
implying the condition $T_0\ll T$. An additional condition $\pi D_\mathrm{f}(1+|M|)\ll T^2|M|$ is imposed on the aperture. Both conditions have their analogs in the spatial imaging which can be found in Ref.~\cite{Tichenor1972}. In the limit of high magnification, $|M|\gg1$, the last condition reads $T_r\ll T$, where $T_r$ is the temporal resolution of the imaging system, defined as \cite{Kolner1994}
\begin{equation}\label{timeres}
T_r = \frac{2\pi D_\mathrm{f}}{T}.
\end{equation}

Now we pass to the full quantum treatment of the field transformation in a temporal imaging system. In distinct contrast to the classical field amplitude, the quantum field operator $\hat{A}_\mathrm{in}\left(\tau'\right)$ is not equal to zero even outside the interval $|\tau'|\le T_0/2$. Therefore, the justification of the approximation $\theta\approx0$ should be modified in the quantum formalism. In the quantum case a condition is imposed on the state of the field rather than the field operator. We demand that the state of the object field at $|\tau'|>T_0/2$ is a vacuum state. Additionally, we demand that the field in the image plane is measured, either by direct or homodyne detection, rather than used for subsequent optical processing. Under these two conditions the operator $\hat{A}_\mathrm{in}\left(\tau'\right)e^{i\theta}$ corresponds to a vacuum field with a shifted phase. Since the phase of the vacuum has no physical significance, it can be omitted, which justifies the approximation $\theta\approx0$ for the first summand in the right hand side of Eq.~\eqref{Arect}. A similar argument applies to the second summand under condition that the field $\hat{B}_\mathrm{in}(\tau')$ is in the vacuum state for all $\tau'$.

Under the above conditions the phase $\theta(\tau,\tau')$ can be neglected and Eq.~\eqref{Arect} can be rewritten in the following simple form
\begin{eqnarray}\label{Arect3}
    \hat{A}_{\mathrm{out}}(\tau) &=& \frac{i}{\sqrt{|M|}}\exp\left(-\frac{i\tau^2}{2|M|D_\mathrm{f}}\right)\\\nonumber
    &\times& \left\{\int_{-\infty}^{\infty} p\left(\tau-\tau'\right) \hat{A}_\mathrm{in}\left(\frac{\tau'}M\right)d\tau'\right.\\\nonumber
    &+& \left.\int_{-\infty}^{\infty} q\left(\tau-\tau'\right) \hat{B}_\mathrm{in}\left(\frac{\tau'}M\right)d\tau'\right\},
\end{eqnarray}
The overall transformation, Eq.~(\ref{Arect3}), consists of three elementary transformations for each input field: (i) scaling of time with the factor $M$, (ii) temporal convolution with a time-invariant transfer function, and (iii) multiplication by a quadratic in time phase factor. In the next section we show that such evolution corresponds to a simple transformation of the spectrum observed in a homodyne measurement with a properly chosen local oscillator.

In the limiting case of infinitely long temporal aperture, $T\rightarrow\infty$, and the conversion efficiency equal to unity, we have $s(\tau)=1$, and, as a consequence, $p(\tau)=\delta(\tau)$, $q(\tau)=0$, wherefrom
\begin{eqnarray}\label{Arect4}
     \hat{A}_{\mathrm{out}}(\tau) &=& \frac{i}{\sqrt{|M|}}\exp\left(-\frac{i\tau^2}{2|M|D_\mathrm{f}}\right) \hat{A}_\mathrm{in}\left(\frac{\tau}M\right),
\end{eqnarray}
which reproduces the result of Ref.~\cite{Patera2015}.

\subsection{Homodyne measurement in the output plane}

Let us assume that the mean field at the input of the temporal imaging system has one spectral component at the frequency $\Omega_0$: $\langle\hat A_\mathrm{s}(\tau,z_{\mathrm{in}})\rangle = E_0e^{-i\Omega_0\tau}$. Then from Eq.~(\ref{Arect4}) we obtain
\begin{eqnarray}\label{Amean}
     \langle\hat A_\mathrm{i}(\tau,z_{\mathrm{out}})\rangle &=& \frac{iE_0}{\sqrt{|M|}}e^{-i\Omega_0\tau/M-\frac{i\tau^2}{2|M|D_\mathrm{f}}}
\end{eqnarray}
that is, the output field has one spectral component at the frequency $\Omega_0/|M|$ and an additional chirp. For successful detection of this component one can employ a homodyne measurement with a chirped local oscillator $E_{LO}(\tau)=\mathcal{E}e^{-i\omega_i\tau-\frac{i\tau^2}{2|M|D_\mathrm{f}}}$. In this case the detector photocurrent gives the quadrature of the operator
\begin{equation}\label{Achirp}
    \hat A_D(\tau) = \hat A_\mathrm{i}(\tau,z_{\mathrm{out}})e^{\frac{i\tau^2}{2|M|D_\mathrm{f}}}.
\end{equation}
Local oscillator shaped in this way allows one to measure the spectrum of the quadratures of $A_D(\tau)$. The required chirp in the local oscillator can be obtained, for instance, by mixing a coherent monochromatic wave at the signal wavelength with the dispersed signal before the time lens \cite{Kolobov1999,Tikan2018}. After having passed through the time lens with the signal, the local oscillator will acquire the same chirp as the latter.

Substituting Eq.~\eqref{Arect3} into Eq.~\eqref{Achirp} and taking Fourier transform of both sides we arrive at the following relation:
\begin{eqnarray}\label{aD}
     \hat a_D(\Omega) &=& \sqrt{|M|}s(D_\mathrm{out}\Omega) \hat a_\mathrm{s}(|M|\Omega,z_{\mathrm{in}})\\\nonumber
     &+& \sqrt{|M|}c'(D_\mathrm{out}\Omega)\hat a_\mathrm{i}(|M|\Omega,z_{\mathrm{in}}),
\end{eqnarray}
where the Fourier components for all fields are defined as
\begin{equation}\label{Fourier}
     \hat a_\mu(\Omega) = \int_{-\infty}^{\infty}\mathrm{d}\tau \mathrm{e}^{\mathrm{i}\tau \Omega} \hat A_\mu(\tau).
\end{equation}

The measured quadrature is $\hat X_D(\Omega)=\hat a_D(\Omega)e^{-i\varphi}+\hat a_D^\dagger(-\Omega)e^{i\varphi}$, where $\varphi$ is the phase of the local oscillator. The spectrum $S_D(\Omega)$ of this quadrature is defined as $\langle \hat X_D^\dagger(\Omega) \hat X_D(\Omega')\rangle = 2\pi S_D(\Omega)\delta(\Omega+\Omega')$ and can be obtained from Eq.~\eqref{aD} as
\begin{equation}\label{spec}
   S_D(\Omega) = |s(D_\mathrm{out}\Omega)|^2 S_{\mathrm{in}}(|M|\Omega) + |c(D_\mathrm{out}\Omega)|^2,
\end{equation}
where $S_{\mathrm{in}}(\Omega)$ of the spectrum of the quadrature $\hat X_{\mathrm{in}}(\Omega)=\hat a_\mathrm{s}(\Omega,z_{\mathrm{in}})e^{-i\varphi} +\hat a_\mathrm{s}^\dagger(-\Omega,z_{\mathrm{in}})e^{i\varphi}$, and we have applied the commutation relation  $[\hat a(\Omega),\hat a^\dagger(\Omega')]=2\pi\delta(\Omega-\Omega')$.

Equation~(\ref{spec}) represents the general rule of the quadrature spectrum transformation in a time-invariant temporal imaging system, and is the main result of the present work. In the next section we illustrate it by an example, where a broadband squeezed vacuum is transformed by a temporal imaging system.

\section{Quantum temporal imaging with broadband squeezed light}

Here we consider an application of the theory developed above to the case where a temporally broadband squeezed state is injected into the input port of a temporal imaging scheme. Such a state of light can be generated in a traveling-wave optical parametric amplifier (OPA) based on a second-order nonlinear crystal~\cite{Kolobov1999}. The broadband squeezing produced by such an OPA is given by a Bogolubov transformation of the photon annihilation operator $\hat{a}_{\mathrm{s}}(\Omega,z)$ from the input ($z=0$) to the output ($z=l$) of the OPA,
\begin{equation}
     \hat{a}_{\mathrm{s}}(\Omega,l)=
     U(\Omega)\hat{a}_{\mathrm{s}}(\Omega,0)+V(\Omega)\hat{a}^{\dag}_{\mathrm{s}}(-\Omega,0). \label{Bogolubov}
\end{equation}
Here $l$ is the length of the OPA crystal, and $U(\Omega)$ and $V(\Omega)$ are the following complex coefficients \cite{Kolobov1999}:
\begin{eqnarray}\label{UV}
U(\Omega) &=& e^{i(k_o(\Omega)-k_l-\Delta(\Omega)/2)l} \\\nonumber
&\times& \left[\cosh(\Gamma l) + \frac{i\Delta(\Omega)}{2\Gamma}\sinh(\Gamma l)\right],\\\nonumber
V(\Omega) &=& e^{i(k_o(\Omega)-k_l-\Delta(\Omega)/2)l} \frac{\sigma}{\Gamma}\sinh(\Gamma l),
\end{eqnarray}
where $k_o(\Omega)$ is the wave vector of the signal in the OPA medium, $k_l=k_o(0)$, $\sigma$ is the coefficient of nonlinear coupling, proportional to the nonlinear susceptibility of the OPA medium and the pump amplitude, $\Delta(\Omega) = k_o(\Omega)+k_o(-\Omega)-2k_l$ is the phase mismatch function, and $\Gamma=\sqrt{|\sigma|^2 - \Delta(\Omega)^2/4}$. For the sake of simplicity we apply a quadratic approximation to the dispersion law of the OPA crystal, similar to Eq.~\eqref{quadratic}, which gives the following approximation for the phase mismatch function:
\begin{equation}
     \Delta(\Omega) \approx \frac2{l}\frac{\Omega^2}{\Omega_c^2},
\end{equation}
where $\Omega_c=(\beta_o^{(2)}l/2)^{-1/2}$ is the characteristic frequency of the squeezed light at the OPA output.

The balanced homodyne photodetection at the output of the OPA returns the normalized to the shot-noise photocurrent noise spectrum $(\delta i)^2_{\mathrm{\Omega}}/\langle i\rangle$, which gives the spectrum of the field quadrature corresponding to the local oscillator phase $\varphi$, and which we shall call spectrum of squeezing and denote $S(\Omega)$. For the unit photodetection efficiency, we can write the spectrum of squeezing as follows:
\begin{equation}
     S(\Omega)=\cos^2\left[\psi(\Omega)-\varphi\right]\mathrm{e}^{2r(\Omega)}
     +\sin^2\left[\psi(\Omega)-\varphi\right]\mathrm{e}^{-2r(\Omega)}, \label{squeezing}
\end{equation}
where $r(\Omega)=\ln \left(|U(\Omega)|+|V(\Omega)|\right)$ is the degree of squeezing and $\psi(\Omega)=\arg[U(\Omega)V(-\Omega)]/2$ is the angle of squeezing at given frequency $\Omega$.  In the examples considered below we put $\varphi=\pi/2$, which corresponds to observation of maximal squeezing at the degeneracy, $\Omega=0$.

Major part of the spectrum of squeezing occupies the frequency band below the frequency $\Omega_q=\Omega_c(|\sigma|^2l^2+\pi^2)^{1/4}$, which is the first zero of $V(\Omega)$, corresponding to the second intersection of $S(\Omega)$ with 1. The light generated by the OPA in this band is squeezed, but its angle of squeezing varies with the frequency $\Omega$ due to the dispersion in the OPA crystal. If the phase $\varphi$ of local oscillator is tuned to the observation of squeezing at the degeneracy, $\Omega=0$, then at the higher frequencies the observed spectrum corresponds to the stretched quadrature component and is above the unity. Observation of squeezing at all frequencies below $\Omega_q$ is possible with the help of quadratic dispersion compensation \cite{Patera2015}. We shall call the frequency $\Omega_q$ the squeezing bandwidth of the broadband squeezed light. It always surpasses the characteristic frequency $\Omega_c$, $\Omega_q>\Omega_c$, which roughly corresponds to the observed squeezing bandwidth without dispersion compensation.

When the imaging condition, Eq.~\eqref{lenseq}, is met, the ideal point-spread function, $p(\tau)=\delta(\tau)$, (with infinitely long pump and unit conversion efficiency) induces a rescaling of the noise spectrum at the output \cite{Patera2015} :
\begin{equation}
   S_{\mathrm{out}}(\Omega)=S_{\mathrm{in}}(|M|\Omega).
\label{noise spectrum perfect}
\end{equation}
This result shows that the squeezing spectrum at the output of the imaging system will be the same as that at the output of the OPA in terms of the scaled frequency $\Omega'=|M|\Omega$ for any magnification factor $M$. This corresponds, in the time domain, to the scaled time $\tau'=\tau/M$. This magnification factor gives us a possibility of matching the quantum correlation time $\tau_q=2\pi/\Omega_q$ of the broadband squeezed light to the response time of the photodetector.

For a non-ideal situation the squeezing spectrum $S_{\mathrm{out}}(\Omega)$ for the output field can be obtained by using the unitary transformation \eqref{aD}. We shall assume that the local oscillator is shaped such that the residual quadratic phase impressed onto the output field by the imaging scheme is compensated. Then from Eqs.~\eqref{spec} and \eqref{unit} the output squeezing spectrum is
\begin{eqnarray}
S_{\mathrm{out}}(\Omega) &=& \eta\,P^2\left(|M-1|\Omega/\Omega_r\right) S_{\mathrm{in}}(|M|\Omega)\\\nonumber
&+& 1-\eta\,P^2\left(|M-1|\Omega/\Omega_r\right).
\label{noise finite aperture}
\end{eqnarray}
where $\eta=|s(0)|^2$ is the conversion efficiency at the center of the pump pulse, and $P(\tau/T)=|s(\tau)/s(0)|$ is the pupil function of a time lens with the temporal aperture $T$. Comparison of this expression with Eq. \eqref{noise spectrum perfect} shows that, as in the case of infinite aperture, the output spectrum is rescaled by factor $|M|^{-1}$. Hence, as discussed above, the output squeezing bandwidth results to be $\Omega_q'=\Omega_q/|M|$. Additional effect of the finite time lens aperture consists in a filtering that is characterized by the bandwidth $\Omega_{r}$, given by the inverse of the resolution time of the imaging system $T_r=2\pi/\Omega_{r}=2\pi D_{\mathrm{f}}/T$.

\begin{figure}[h!]
    \centering
    \includegraphics[width=\columnwidth]{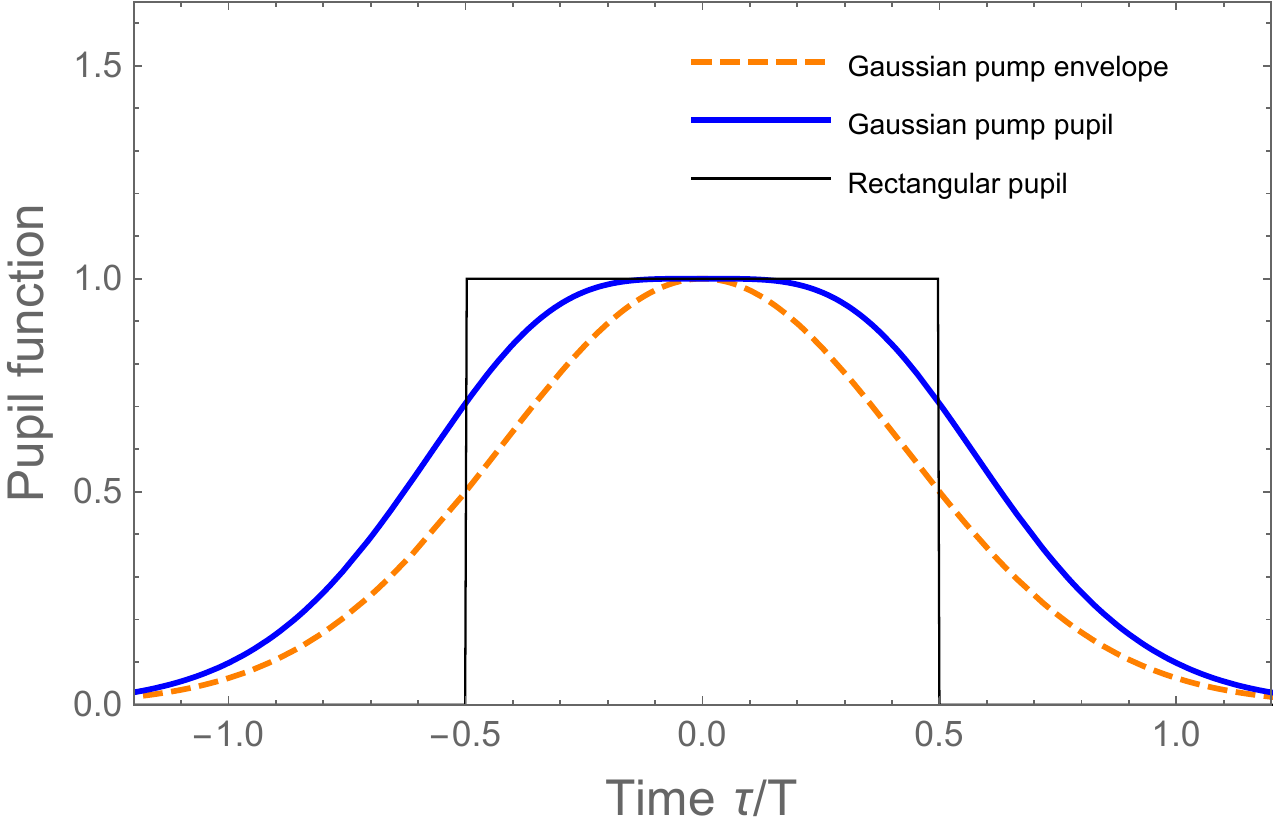}
  \caption{Pupil functions $P(\tau/T)$ for an SFG-based time lens: one corresponding to Gaussian pump envelope (thick blue line) and one corresponding to rectangular pump envelope (thin black line). Normalized Gaussian pump envelope with unit full width at half maximum is shown by dashed orange line.}
  \label{fig:Pupils}
\end{figure}

Below we illustrate the transformation of the spectrum of squeezing in an SFG-based temporal imaging system for a Gaussian pump. We define a Gaussian pump envelope with unit full-width at half maximum, $A_{\mathrm{p}}(\tau) = A_{\mathrm{p}}(0)\exp\{-4\ln2(\tau/T)^2\}$. We assume for simplicity the ideal case of unit conversion efficiency, $\eta=1$. In this case the pupil function is
\begin{align}
    P(\tau/T)&=\sin\left(\frac\pi2\exp\{-4\ln2(\tau/T)^2\}\right),
\label{pupilG}
\end{align}
and it corresponds to using a dispersed Gaussian pulse for pumping the SFG crystal. Another interesting example is a rectangular pump envelope $A_{\mathrm{p}}(\tau) = A_{\mathrm{p}}(0)\,\mathrm{rect}(\tau/T)$, which results in a rectangular pupil function $P(\tau/T)=\mathrm{rect}(\tau/T)$. Such a pulse corresponds exactly to the aperture of a conventional lens and can be produced with certain precision by cutting out the central part of a dispersed Gaussian pulse by means of a pulse shaper. The examples of transformation of spectra of squeezing for this type of pupil function can be found in Ref.~\cite{PateraSPIE2018}. Both pupil shapes are shown in Fig.~\ref{fig:Pupils}.

\begin{figure}[h!]
    \centering
    \includegraphics[width=\columnwidth]{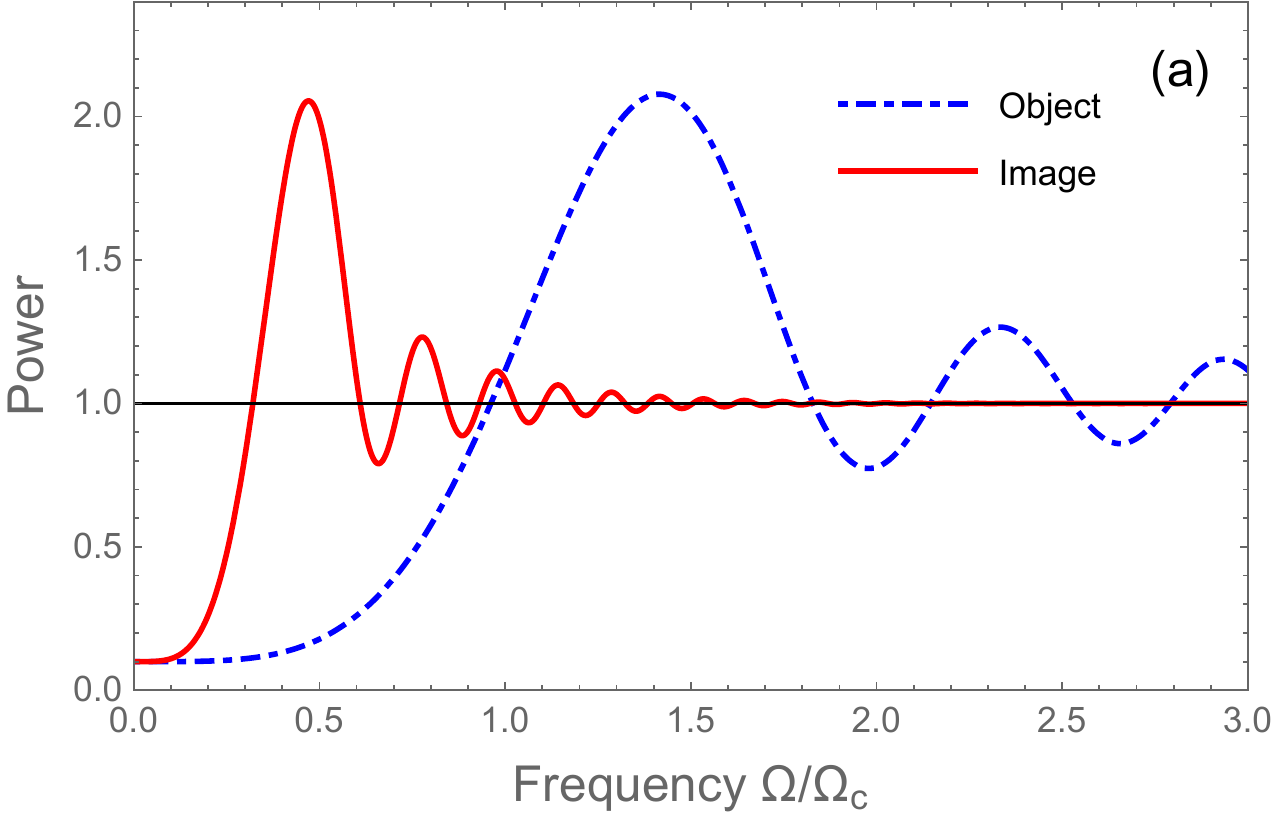}
    \vskip 0.3cm
    \includegraphics[width=\columnwidth]{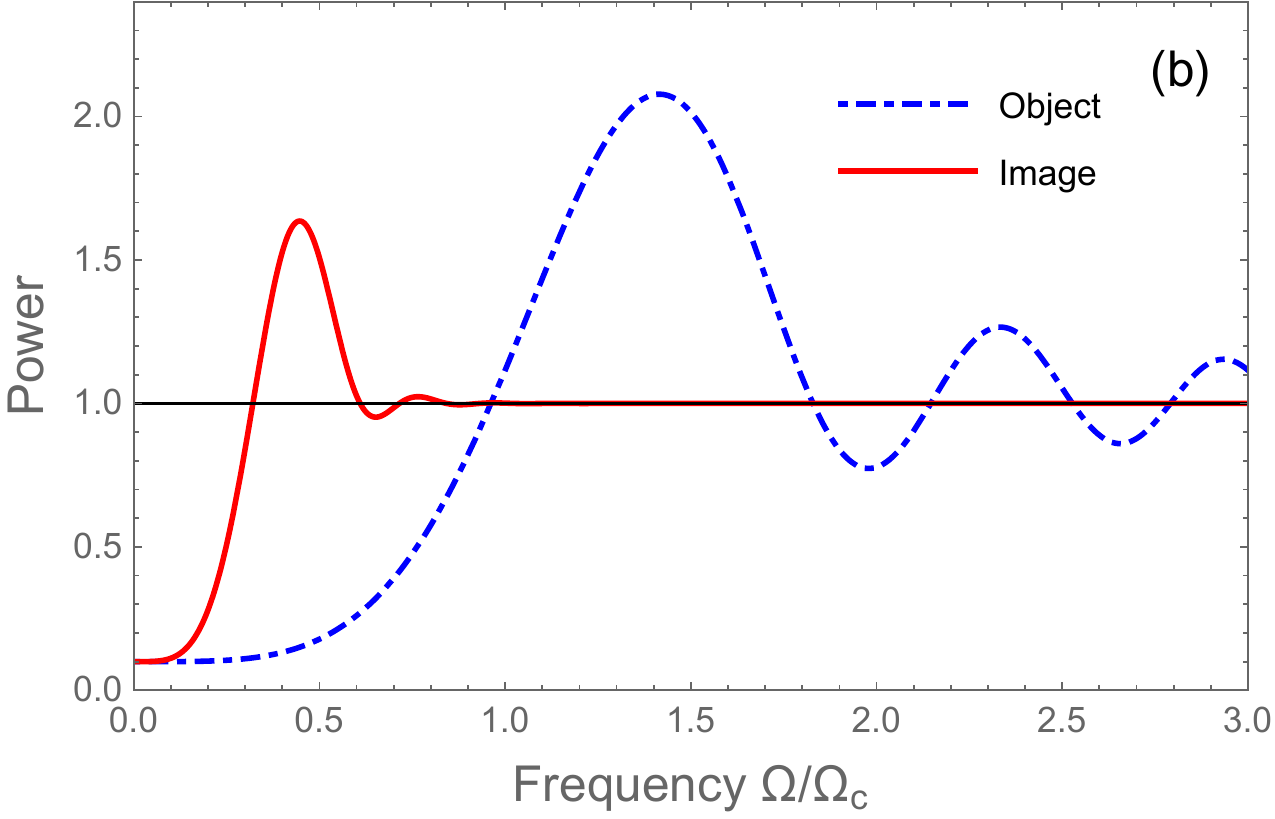}
  \caption{Transformation of the spectrum of squeezing in an SFG-based temporal imaging system with magnification $M=-3$. The spectrum of squeezing of the image field (solid red line) is a scaled and filtered version of the spectrum of squeezing of the object field (dot-dashed blue line). All spectra are normalized to the shot noise level (horizontal thin black line). Temporal aperture of the time lens is chosen so that (a) $\Omega_r=10\Omega_c$, where the filtering is unnoticeable, and (b) $\Omega_r=4\Omega_c$, where the filtering is noticeable. }
  \label{fig:GaussPupil}
\end{figure}

The spectra of squeezing before and after the temporal imaging system with Gaussian pump are shown in Fig.~\ref{fig:GaussPupil} for magnification $M=-3$ and the conversion efficiency $\eta=1$. The squeezing is 10 dB at $\Omega=0$, corresponding to $r(0)=1.15$.

As follows from Fig.~\ref{fig:GaussPupil}, the spectrum of squeezing of the image field is a scaled and filtered version of the spectrum of squeezing of the object field. In the regime of magnification, $|M|>1$, the spectrum is compressed. In Fig.~\ref{fig:GaussPupil}a the filter bandwidth is so wide compared to the squeezing bandwidth $\Omega_q=1.8\Omega_c$, that its effect is negligible, and the transformation can be described by Eq.~\eqref{noise spectrum perfect}. In Fig.~\ref{fig:GaussPupil}b the effect of filtering is visible in the decreased first maximum in the image spectrum compared to the first maximum of the object spectrum.

The filtering can be described as cutting off the image spectrum above the cutoff frequency
\begin{equation}
\Omega_\mathrm{cutoff} = \frac{\Omega_r}{2|M-1|} =  \frac{T}{2|M-1|D_{\mathrm{f}}} .
\end{equation}
The cutoff frequency of a temporal imaging system is proportional to the temporal aperture $T$ and can be made sufficiently large by increasing the latter. In the limit of high magnification, $|M|\gg1$, the filtering bandwidth scales as $|M|^{-1}$, similarly to the correlation time of the system. Thus, if the object squeezing bandwidth $\Omega_q$ is below $\Omega_r/2$, then the image squeezing bandwidth $\Omega_q'$ is below $\Omega_\mathrm{cutoff}$, and the important non-classical part of the squeezing spectrum does not suffer from the filtering. In the opposite limit of high reduction factor, $|M|\ll1$, the situation is more complicated. Indeed, the cutoff frequency in this limit is $\Omega_r/2$ for any $|M|$, while the image squeezing bandwidth is increased $|M|^{-1}$ times. As consequence, for any given time lens aperture there is such a reduction factor $|M|<1$, for which the squeezing in the image field is significantly modified by filtering.

\begin{figure}[h!]
    \centering
    \includegraphics[width=\columnwidth]{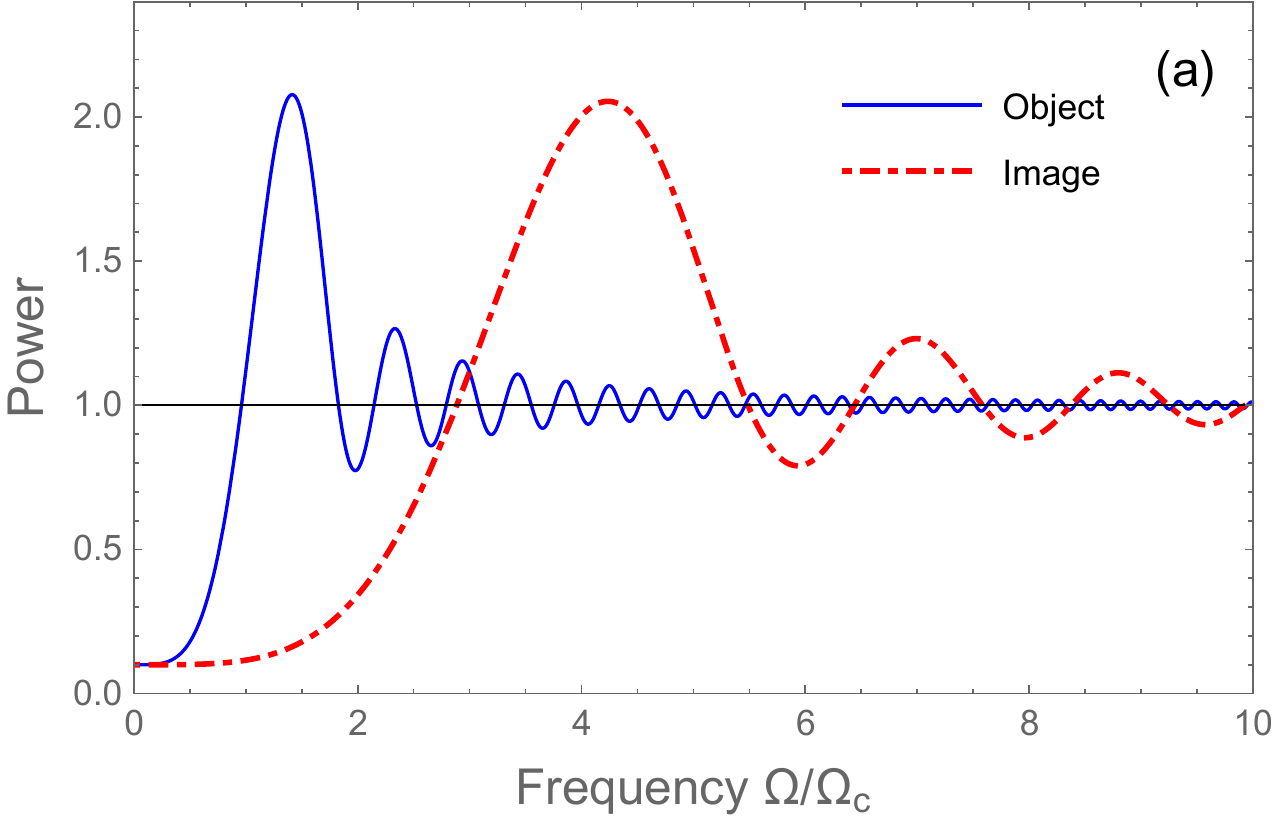}
    \vskip 0.3cm
    \includegraphics[width=\columnwidth]{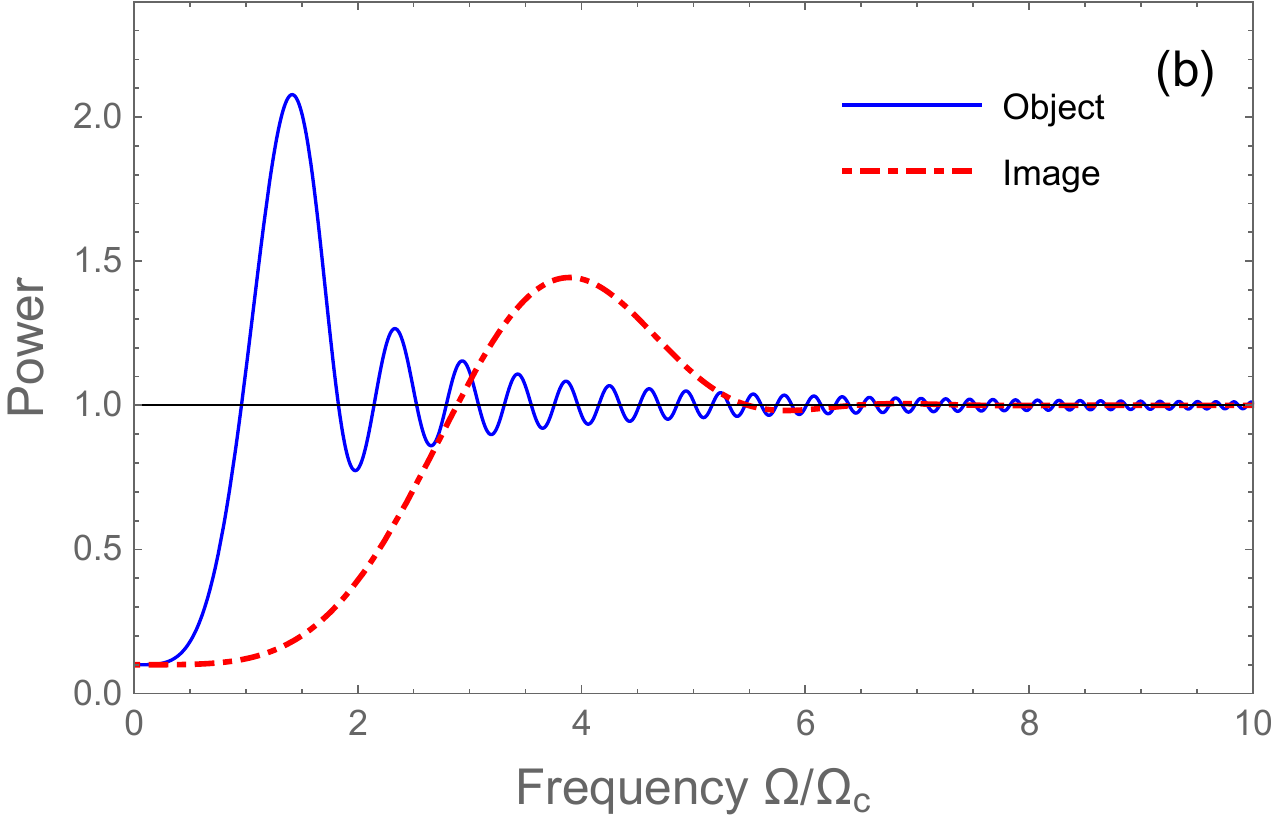}
  \caption{Transformation of the spectrum of squeezing in an SFG-based temporal imaging system with magnification $M=-1/3$. The spectrum of squeezing of the image field (dot-dashed red line) is a scaled and filtered version of the spectrum of squeezing of the object field (solid blue line). All spectra are normalized to the shot noise level (horizontal thin black line). Temporal aperture of the time lens is chosen so that (a) $\Omega_r=30\Omega_c$, where the filtering is unnoticeable, and (b) $\Omega_r=10\Omega_c$, where the filtering is well noticeable. }
  \label{fig:GaussPupil-M03}
\end{figure}

This situation is illustrated in Fig.~\ref{fig:GaussPupil-M03} for reduction factor $M=-1/3$ and other parameters as in Fig.~\ref{fig:GaussPupil}.

In Fig.~\ref{fig:GaussPupil-M03}a the cutoff frequency $\Omega_\mathrm{cutoff}=11.25\Omega_c$ is so high compared to the squeezing bandwidth $\Omega_q'=5.4\Omega_c$ that its effect is negligible, and the transformation can be described by Eq.~\eqref{noise spectrum perfect}. In Fig.~\ref{fig:GaussPupil-M03}b the effect of filtering is visible in the decreased first maximum in the image spectrum and complete suppression of oscillations beyond it, which are well above $\Omega_\mathrm{cutoff}=3.75\Omega_c$.

\begin{figure}[h!]
    \centering
    \includegraphics[width=\columnwidth]{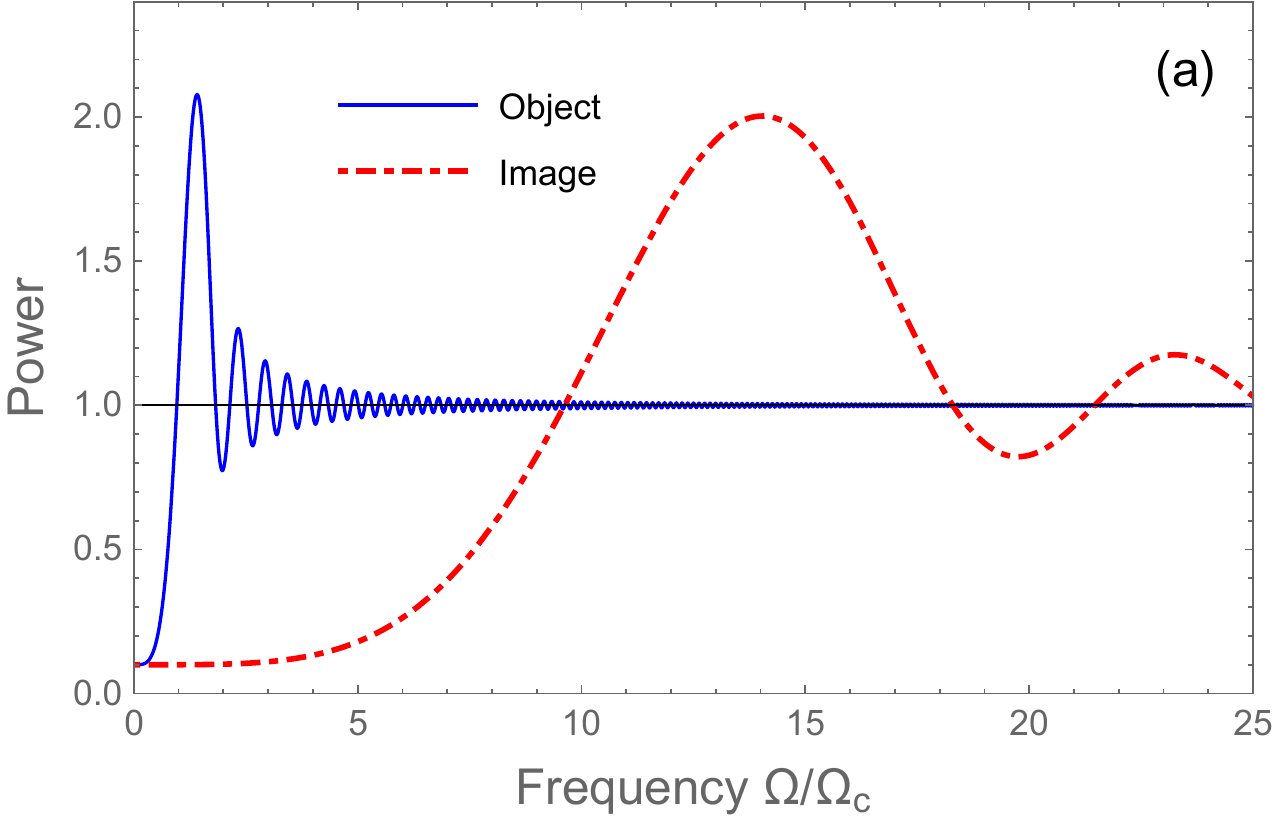}
    \vskip 0.3cm
    \includegraphics[width=\columnwidth]{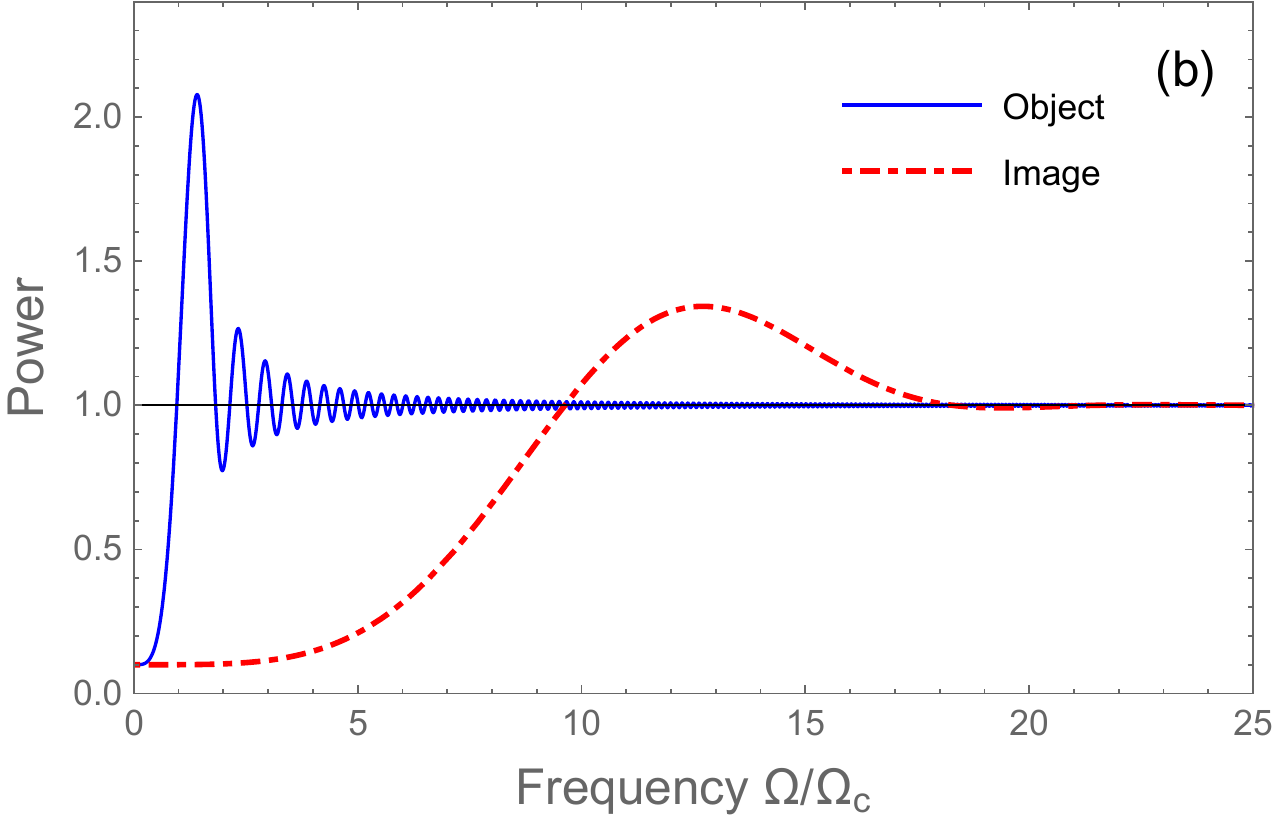}
    \vskip 0.3cm
    \includegraphics[width=\columnwidth]{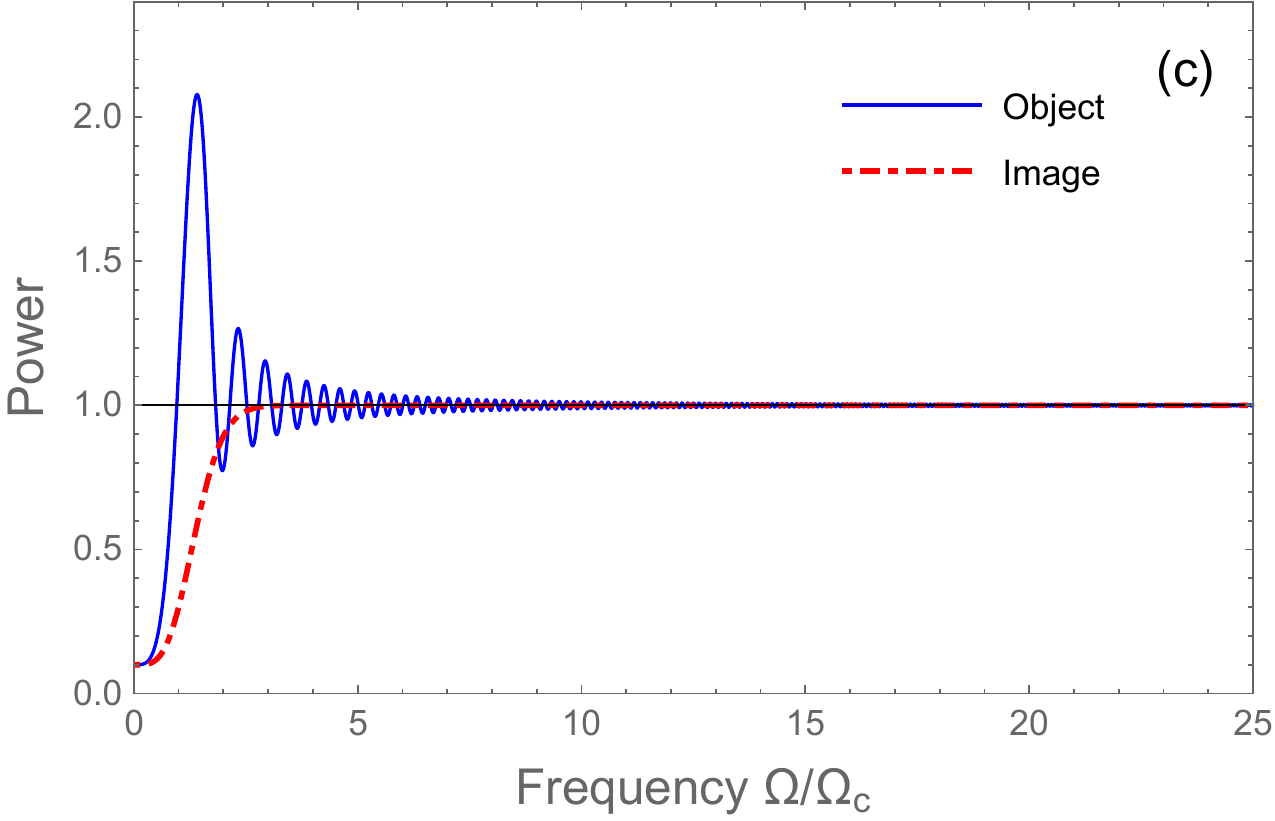}
  \caption{Transformation of the spectrum of squeezing in an SFG-based temporal imaging system with magnification $M=-0.1$. The spectrum of squeezing of the image field (dot-dashed red line) is a scaled and filtered version of the spectrum of squeezing of the object field (solid blue line). All spectra are normalized to the shot noise level (horizontal thin black line). Temporal aperture of the time lens is chosen so that (a) $\Omega_r=60\Omega_c$, (b) $\Omega_r=25\Omega_c$, and (c) $\Omega_r=3\Omega_c$. In the last case the effect of filtering is detrimental for the squeezing bandwidth.}
  \label{fig:GaussPupil-M01}
\end{figure}

In Fig.~\ref{fig:GaussPupil-M01} we show the squeezing spectra for even larger reduction factor $M=-0.1$ and the other parameters as before.

In Fig.~\ref{fig:GaussPupil-M01}a the cutoff frequency $\Omega_\mathrm{cutoff}=30\Omega_c$ is so high compared to the squeezing bandwidth $\Omega_q'=18\Omega_c$, that the filtering effect is negligible, and the transformation can be described by Eq.~\eqref{noise spectrum perfect}. In Fig.~\ref{fig:GaussPupil-M01}b the effect of filtering is visible in the decreased first maximum in the image spectrum. Note that in this case the filter bandwidth is one order of magnitude higher than the object bandwidth $\Omega_q=1.8\Omega_c$. However, this is not sufficient for preserving the quantum features of the light above the cutoff frequency $\Omega_\mathrm{cutoff}=12.5\Omega_c$. In Fig.~\ref{fig:GaussPupil-M01}c only a small part of the stretched squeezing spectrum remains unchanged, while the rest of the squeezing has been lost. Because of the strong filtering, the image squeezing bandwidth $\Omega_\mathrm{cutoff}=1.5\Omega_c$ is lower than that of the object, even if the spectrum was intended to be ``stretched''.

With these examples we have demonstrated that the resolution time $T_r$ is a key parameter of the time lens, limiting the performance of a temporal imaging system, especially at the high reduction factor, $|M|\ll1$ . Let us clarify the practical condition for decreasing $T_r$. When the SFG pump is obtained by dispersing a Fourier-limited Gaussian pulse of duration $\tau_{\mathrm{p}}$ through a medium with a GDD equal to $-D_{\mathrm{f}}$, the chirped pump pulse has a duration $T=2\pi D_{\mathrm{f}}/\tau_{\mathrm{p}}$. In this case the temporal resolution of the time lens is given by the duration of the initial pulse $T_r=\tau_{\mathrm{p}}$. Therefore, shorter pump pulses produce time lenses with better resolution.

\section{Conclusions}

We have extended the quantum theory of temporal imaging formulated in our previous works in order to take into account the effect of finite duration of a chirped pump pulse used in a nonlinear SFG time lens. This problem has been considered in the classical temporal imaging theory \cite{Kolner1994,Bennett1999,Bennett2000a,Bennett2000b,Bennett2001}, where it was demonstrated that this effect can be described by a finite temporal pupil function and impulse response function imposing a finite resolution of the imaging scheme. We have demonstrated that in the framework of the quantum theory such a classical description is insufficient because it contradicts to the conservation of the commutation relations of the quantum field operators at the output of the imaging scheme. In order to restore these commutation relations, we have found the missing in classical theory part of the electromagnetic field describing the contribution of the vacuum fluctuations into the output of the imaging scheme. We have also determined a second impulse response function describing the transformation of these vacuum fluctuation from the input to the output of the imaging scheme. These vacuum fluctuations are neglected in the classical temporal imaging but have to be taken into account for imaging of nonclassical states of light such as squeezed, entangled, or sub-Poissinian states. As an example of application of our theory, we have considered illumination of our imaging scheme by broadband squeezed light. We have demonstrated that the squeezing spectrum at the output of the scheme is a rescaled and filtered copy of the input squeezing spectrum. We have formulated the criteria for preservation of squeezing at the output of the scheme for different values of the magnification and the bandwidth of the squeezing spectrum vs the width of the temporal pupil function. Our results can have numerous application for temporal imaging of nonclassical states of light in quantum optics and quantum information.

\section*{Acknowledgments}

This work was supported by the European Union's Horizon 2020 research and innovation programme under grant agreement No 665148 (QCUMbER).

\end{document}